\documentclass[conference]{IEEEtran}

\IEEEoverridecommandlockouts
\usepackage{etex}

\usepackage{cite}
\usepackage[numbers,sort&compress]{natbib}

\usepackage{amsmath,amssymb,amsfonts}
\usepackage{graphicx}
\usepackage{subcaption}
\usepackage{textcomp}
\usepackage{amsmath}
\usepackage{xcolor}
\usepackage{caption}
\usepackage{float}

\usepackage{algorithm} 
\usepackage{algorithmic}  
\usepackage[algo2e]{algorithm2e} 
\usepackage{url}
\usepackage{array}
\usepackage{booktabs}
\usepackage{hhline}
\graphicspath{ {./fig/} }
\usepackage[section]{placeins}
\usepackage{mathtools}
\usepackage{multirow}

\graphicspath{ {./figures/} }
\def\BibTeX{{\rm B\kern-.05em{\sc i\kern-.025em b}\kern-.08em
    T\kern-.2em\lower.7ex\hbox{E}\kern-.125emX}}

\begin{document}

\title{LLM-supported 3D Modeling Tool for Radio Radiance Field Reconstruction \thanks{This project is partially supported by the U.S. National Science Foundation under Grants ECCS-2336234 and CNS-2236449.}}

\author{\IEEEauthorblockN{Chengling Xu\IEEEauthorrefmark{1}, Huiwen Zhang\IEEEauthorrefmark{1}, Haijian Sun\IEEEauthorrefmark{2} and Feng Ye\IEEEauthorrefmark{1}}
\IEEEauthorblockA{\IEEEauthorrefmark{1}Department of Electrical and Computer Engineering, University of Wisconsin-Madison, Madison, WI, USA\\
\IEEEauthorrefmark{2}School of Electrical and Computer Engineering
University of Georgia, Athens, GA, USA\\
Emails: \IEEEauthorrefmark{1}\{cxu338, hzhang2279, feng.ye\}@wisc.edu,  \IEEEauthorrefmark{2}hsun@uga.edu}
}

\maketitle

\begin{abstract}
Accurate channel estimation is essential for massive multiple-input multiple-output (MIMO) technologies in next-generation wireless communications. Recently, the radio radiance field (RRF) has emerged as a promising approach for wireless channel modeling, offering a comprehensive spatial representation of channels based on environmental geometry. State-of-the-art RRF reconstruction methods, such as RF-3DGS, can render channel parameters, including gain, angle of arrival, angle of departure, and delay, within milliseconds. However, creating the required 3D environment typically demands precise measurements and advanced computer vision techniques, limiting accessibility. This paper introduces a locally deployable tool that simplifies 3D environment creation for RRF reconstruction. The system combines finetuned language models, generative 3D modeling frameworks, and Blender integration to enable intuitive, chat-based scene design. Specifically, T5-mini is finetuned for parsing user commands, while all-MiniLM-L6-v2 supports semantic retrieval from a local object library. For model generation, LLaMA-Mesh provides fast mesh creation, and Shap-E delivers high-quality outputs. A custom Blender export plugin ensures compatibility with the RF-3DGS pipeline. We demonstrate the tool by constructing 3D models of the NIST lobby and the UW–Madison wireless lab, followed by corresponding RRF reconstructions. This approach significantly reduces modeling complexity, enhancing the usability of RRF for wireless research and spectrum planning.
\end{abstract}

\section{Introduction}\label{sec:introduction}

Current and next-generation wireless communication systems rely heavily on multiple-input multiple-output (MIMO) and massive MIMO technologies, including beamforming, delay-alignment modulation, and reconfigurable intelligent surfaces. Accurate channel information is essential for optimizing these technologies. Recently, the radio radiance field (RRF) has emerged as a promising approach for spatial representation of wireless channels~\cite{Zhao_2023,10589576,zhang2025rf3dgswirelesschannelmodeling}. Unlike traditional channel estimation and modeling techniques, an RRF characterizes how radio waves propagate from every point in a 3D space toward the entire spherical domain. This representation captures delay, angle of arrival, angle of departure, and polarization in addition to conventional channel gain.
However, reconstructing an RRF, such as with RF-3DGS~\cite{zhang2025rf3dgswirelesschannelmodeling},
requires geometric information about a bounded environment, typically obtained through physical measurements and advanced computer vision techniques. Existing 3D reconstruction tools (e.g., Blender) are not directly compatible with current RRF pipelines, creating a significant barrier for researchers.

\begin{figure}[t]
    \centering
    \includegraphics[width=.99\linewidth]{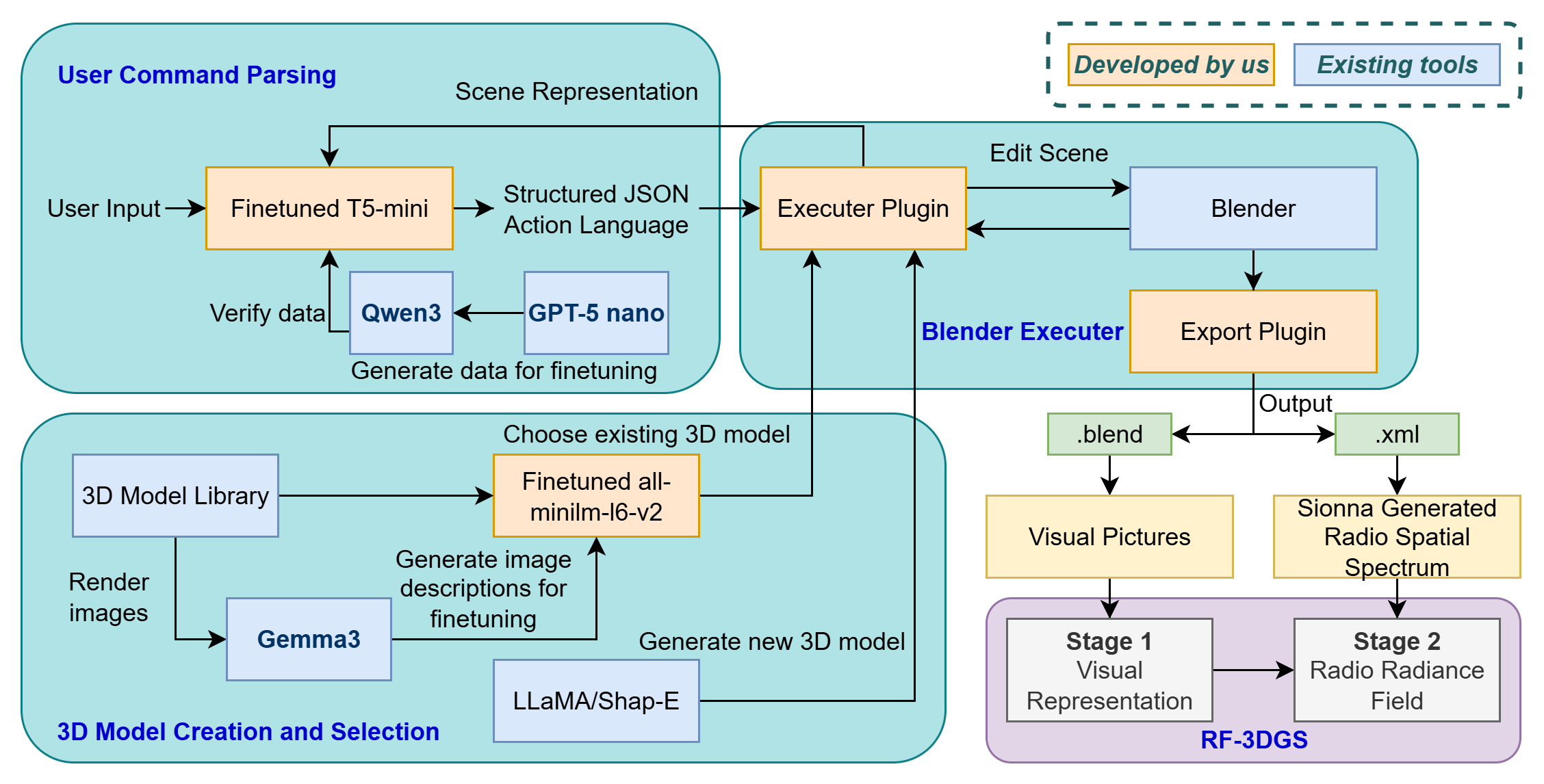}
    \caption{Overview of the LLM-supported 3D modeling tool.}
    \label{fig:structure}
    \vspace{-6mm}
\end{figure}

To address this challenge, we propose an intuitive tool that leverages multiple locally deployed large language models (LLMs) to generate 3D environments compatible with RF-3DGS. Our system enables users to create and manipulate complex 3D models through a simple chat interface, with support for sharing and reusing environments within the research community. While users can directly employ a general-purpose LLM within our framework, we focus on a local deployment in is work for better privacy, latency, and resource efficiency. As illustrated in Fig.~\ref{fig:structure}, the tool consists of three main components: natural language parsing, 3D model creation/selection, and output interfacing.
First, natural language commands are converted into a structured, machine-readable format for 3D applications. This is achieved by combining a general-purpose LLM for intent understanding with a finetuned T5-mini model~\cite{t5_efficient} that translates commands into structured JSON actions. Second, a custom Blender executor plugin applies these actions to edit the scene. Users can create new 3D objects using local generative models such as LLaMA-Mesh~\cite{wang_llama-mesh_2024} or Shap-E~\cite{shap_e}, or search an existing library via a finetuned all-minilm-l6-v2 model~\cite{all_minilm_l6_v2}. This design promotes flexibility and collaboration. Finally, a customized export plugin ensures seamless integration with RF-3DGS by supporting both visual and RF model reconstructions.

We demonstrate the proposed tool by constructing two complex indoor environments: the National Institute of Standards and Technology (NIST) lobby~\cite{10501217} and the wireless lab at the University of Wisconsin-Madison (UW-Madison). Natural language instructions specifying room structure, object dimensions, materials, and fine-grained scene manipulations are provided to the chatbot. The resulting environments are then used for RRF reconstruction with RF-3DGS. Experimental results show that scenes generated through our system are highly comparable to those built using traditional manual methods, highlighting its potential for practical applications.
The remainder of this paper is organized as follows: Section~\ref{sec:llm_parsing} describes the natural language parsing approach using LLMs; Section~\ref{sec:3D modeling} details the 3D modeling process with local generative AI and library integration; Section~\ref{sec:blender_plugin} introduces the customized Blender plugin for RRF pipeline compatibility; and Section~\ref{sec:conclusion} concludes the paper.

\section{A Local LLM for User Command Parsing}

While users can directly employ a general-purpose LLM within our framework for natural language parsing, it introduces heavy computational and storage overhead. To address this problem and inadequate ability of smaller models without adaptation, we present an LLM distillation approach of finetuning a small sequence-to-sequence~\cite{seq2seq} model with synthetic data generated by LLM.

\subsection{Direct Parsing with LLM}\label{sec:llm_parsing}

The primary objective of parsing is to translate a user’s natural language command into a JSON array of action objects, where each object represents a discrete operation in the 3D scene. This structured format ensures deterministic interpretation and execution. For example, given the command ``Add a nightstand, then put a lamp on it'', the second action depends on the first, so actions must be executed sequentially in array order.
Each action object includes an action\_type field, which must match one of 16 predefined values (e.g., create\_object\_relative, move\_object\_offset, change\_object\_material). Additional fields depend on the action type: \textbf{Creation Actions}(``create\_object\_absolute'', ``create\_object\_relative'') require fields like ``object\_type'' (e.g., ``chair''), quantity, and a ``local\_id'' (e.g., ``1''). 
 The local\_id enables subsequent actions to reference newly created objects using the \# prefix.
\textbf{Modification Actions} (move, rotate, resize, delete, change material) require an ``object\_name'' to identify the target, which may refer to an existing scene object or a local ID. Parameters specify details such as position (x, y, z), rotation angle, size, material, or scale factor.
\textbf{Room Setup Action} (``setup\_room'') initializes the basic room structure, assumed to be a cube, and requires room size parameters
This schema is designed to comprehensively support essential operations for constructing 3D scenes in wireless communication research. More actions can be added straightforwardly if deemed necessary.

Achieving accurate parsing with a general-purpose LLM depends on a rigorous prompt, dynamically constructed for each user command. The prompt includes:
\begin{itemize}
    \item Role definition: A directive specifying the model’s role as an AI assistant for 3D applications and its task of converting user input into a JSON list of actions.

    \item Scene Context: A list of existing objects, valid materials, directions, and room elements. Object names (e.g., ``chair.001'') are descriptive to aid interpretation.
    
    \item Schema Definition: Explicit JSON output rules to enforce structural correctness.
    
    \item Few-Shot Examples: High-quality examples illustrating multi-step commands, relative positioning, and object references.

    \item Input Output Format: The user command appears at the end of the prompt, and the output must be enclosed in a JSON code block.
\end{itemize}
This structured approach constrains the LLM’s creativity and ensures syntactically and semantically valid JSON outputs.

\subsection{Distill a Local Model With Synthetic Data}\label{sec:distill}

Decoder-only LLMs excel at generating fluent text but often introduce syntax errors and hallucinated fields when producing structured outputs like JSON. Moreover, large models incur significant computational overhead and latency, making them unsuitable for real-time interactive applications. To address these challenges, we adopt a model distillation strategy: First, use a powerful LLM to generate a synthetic, high-quality dataset of natural language commands and corresponding JSON outputs; Second, finetune a smaller, efficient model, specifically Google’s T5~\cite{2020t5}, on this dataset.
Common decoder-only LLMs excel at creative tasks like generating fluent text by sampling the next word from a probability distribution, but their creativity is a liability for structured data with strict grammar like JSON. It often introduces syntax errors and hallucinated fields. 
Although a large language model is a powerful tool for natural language parsing, it introduces significant computational overhead and latency, making it unsuitable for a real-time and interactive application. To address this, we adopted a model distillation approach. We used a powerful LLM to generate a synthetic high-quality dataset, which will then be used to finetune a much smaller and more efficient model, specifically Google's T5~\cite{2020t5}. This finetuned model can perform the parsing task with much lower computational cost and latency.
T5’s encoder-decoder architecture is well-suited for tasks requiring strong alignment between input and output, such as translating natural language into structured JSON. This design ensures deterministic, stable results. We also apply beam search during decoding to improve accuracy.
Synthetic data generation offers several advantages over manual annotation: it is faster, covers edge cases, and allows controlled complexity. By leveraging a capable LLM for data generation, we ensure comprehensive coverage and minimal human effort.

To ensure high quality dataset for model distillation, the creation pipeline consists of two main stages: data generation and multi-stage validation. In the first stage, a generator LLM is prompted with detailed instructions, schema definitions, and parameters that control sample diversity. These parameters specify the number of samples, the complexity of the scene (ranging from none to many objects), and the complexity of user commands (simple, medium, or complex). To further enhance diversity, a random required action is added to each prompt. The generator then produces samples that include a natural language command, a list of existing objects in the scene, and the corresponding structured JSON actions. The second stage ensures the quality and correctness of the generated data through a rigorous validation process. Initially, each sample is checked for valid JSON syntax. Next, a rule-based validator corrects common errors such as inconsistent local ID sequencing and verifies compliance with schema rules, including action types, field names, and data types (e.g., ensuring quantity is an integer). Duplicate samples are then removed to maintain dataset uniqueness. Finally, semantic validation is performed using a reasoning-capable LLM (Qwen3-8B), which evaluates whether the JSON output logically interprets the input command according to the defined rules. This step catches subtle semantic errors and hallucinations that rule-based checks might miss. Only samples that pass all validation stages are included in the final dataset for finetuning, ensuring high-quality training data for the distilled model.

\section{3D Model Creation and Selection}\label{sec:3D modeling}

Our tool allows users to either create a new 3D model or select one from an existing library. Model creation is powered by locally deployed Shap-E~\cite{shap_e} and LLaMA-Mesh~\cite{wang_llama-mesh_2024}, which provide complementary generation capabilities. Shap-E, a diffusion-based model, produces high-quality 3D models from text prompts but operates relatively slowly, making it ideal when visual fidelity is the priority. In contrast, LLaMA-Mesh represents 3D meshes as text tokens using a large language model approach, enabling much faster generation at the cost of lower detail, which makes it better suited for interactive applications. Users can choose Shap-E for superior quality or LLaMA-Mesh for rapid content creation, depending on their requirements for accuracy versus latency.

Alternatively, users can retrieve 3D objects from a local library through a semantic search system that matches descriptive language (e.g., ``a vase with a wide base'') rather than requiring specific identifiers (e.g., ``vase\_0501''). This system ensures that retrieved objects align with the user’s intended visual characteristics and operates in two stages: data preparation and real-time inference. During data preparation, we adopted ModelNet-40~\cite{wu20153d} as the base asset library. For each object, two images were rendered from different viewpoints, and a multi-modal LLM (Gemma3:12B) generated three textual descriptions per image. After removing duplicates, each object had up to six distinct descriptions, forming a rich semantic representation. To overcome the limitations of standard embedding models, which cluster objects by category rather than instance-level differences, we finetuned all-MiniLM-L6-v2 using a contrastive learning approach with triplet loss. This training strategy ensures that descriptions of the same object remain close in embedding space while those of different objects are pushed apart. After finetuning, all descriptions were converted into vector embeddings and indexed using FAISS~\cite{douze2024faiss} for efficient similarity search. During inference, the user’s descriptive phrase is embedded and matched against the stored vectors to retrieve the most semantically similar object from the library.

\section{Custom Blender Executor Plugin}\label{sec:blender_plugin}

\subsection{The Executer Plugin}

To connect the backend Blender, an executor plugin is created as a server, which listens to requests and executing the expected actions. The transferred data is in JSON format through HTTP requests. This plugin has two main functions: getting scene status and modifying the scene.

\begin{figure}[ht!]
    \begin{subfigure}[t]{.47\linewidth}    
    \centering
    \includegraphics[width=0.99\linewidth]{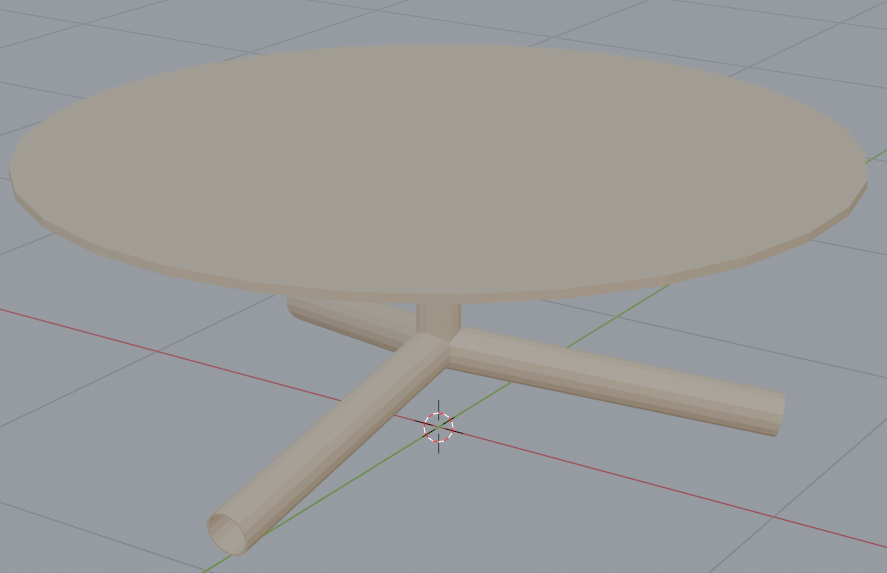}
        \caption{Instruction 1: Create a table in the center of the room}\label{fig:interface_chat1}
    \end{subfigure}
    \hfill
    \begin{subfigure}[t]{.47\linewidth}    
    \centering
    \includegraphics[width=0.99\linewidth]{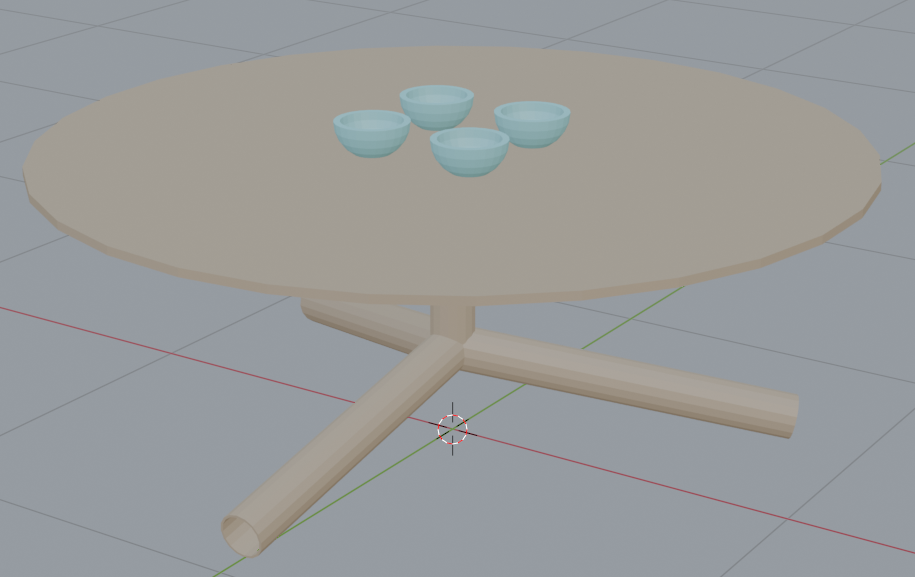}
        \caption{Instruction2: Put 4 bowls on the table}\label{fig:interface_chat2}
    \end{subfigure}
     \caption{A demonstration of 3D objects creation using the chat-based interface.}
    \label{fig:interface_interaction}
    \vspace{-4mm}
\end{figure}

The ``get\_scene'' API returns all necessary scene information. This data includes room dimensions and details of scene objects. Each object is represented by not only standard attributes like location and rotation but also crucial geometric properties such as its geometric center, bounding box coordinates, and overall size for better spatial relation understanding and object placement. As illustrated in Fig.~\ref{fig:interface_chat1}, given a command \emph{``Create a wood table in the center of the room''}, the system confirms the acquisition of the asset, and transmits the object to Blender. 

The second function is to execute modifications to the scene. It receives action requests in series. The abstract parameters have been resolved into concrete operations. For instance, the description for object creation is replaced with a specific filename. Besides, the reference objects using local ID is replaced with actual names. Then relative placements are further calculated into absolute position coordinates. Then the plugin executes each action sequentially by calling the appropriate Blender API function. As shown in Fig.~\ref{fig:interface_chat2},  given a follow-up command \emph{``Create 4 bowls on the table,''}  the system interprets the instruction and generates the corresponding 3D elements within the established scene. 

\subsection{The Blender Export Plugin}

To integrate 3D environments created within Blender into the RF-3DGS pipeline, the output files need to be compatible with an open-source ray tracing simulation tool, Sionna~\cite{sionna}. An existing solution is the Blender plugin called \emph{mitsuba-blender} \cite{mitsuba_blender}. However, this plugin is not available after build of Blender version 4.3 and is not stable on earlier builds either. To address this issue, a custom Blender export plugin was developed. It converts all mesh objects from the Blender scene into individual Polygon File Format (PLY) files and generates a single Extensible Markup Language (XML) file that describes the scene structure and material properties of each object. The plugin only chooses visible mesh objects by using the condition \texttt{obj.type == `MESH' and not obj.hide\_viewport} to export. It also automatically deals with the Sionna built-in material prefix ``itu\_’’ and material name suffix (e.g., ``.001’’) due to object copy. Besides, it applies the common colors of each material for better visual effects.
The main XML file is then written. It begins with a root \texttt{<scene>} tag and everything will be in the root tag. 
Following the header, each material is defined in a \texttt{<bsdf>} (Bidirectional Scattering Distribution Function) tag. These BSDF definitions represent the material properties like type and color. In this implementation, each material is defined as a simple ``twosided’’ BSDF with the pre-defined diffuse RGB color that is expressed by 0-1 decimals.
After defining all materials, the plugin proceeds to export each filtered mesh object as a PLY file in the created ``meshes’’ folder. A standard PLY file needs the information of the vertices’ positions, normal, and indices for faces to represent a mesh. To avoid using extra Python library, we manually constructed the file with the object's mesh data from Blender's bmesh module. For each mesh object, a corresponding \texttt{<shape>} entry in the XML file is needed. This tag defines the relative filename and the material information. The ID and name attributes of the shape are derived from the Blender object's name. 
Finally, after processing all relevant objects, the XML file records all necessary information along with the specified corresponding mesh files. This results in a folder containing of scene that is ready for loading into the RF-3DGS environment.

\section{Implementation and Evaluation Results}\label{sec:eva}

To evaluate the efficacy and practical application of our 3D scene generation tool, visual and RRF models of the NIST lobby and the wireless lab at UW-Madison are generated. Note that the NIST lobby has been physically measured with point cloud data available, while the wireless lab at UW-Madison has no such data.

\subsection{Performance Evaluation on User Command Parsing}

To identify the most effective generator LLM, we evaluated ten models, including variants of Qwen3~\cite{yang2025qwen3}, Gemma3~\cite{team2025gemma}, Deepseek (DS)-R1~\cite{guo2025deepseek}, Llama 3.1~\cite{grattafiori2024llama}, and Phi-4~\cite{abdin2024phi}. Each model was tasked with generating five samples per trial across 100 trials. Performance was assessed using five key metrics decribed below:

\begin{itemize}
    \item \textbf{JSON validity rate}: The percentage of outputs that were syntactically valid JSON lists.

    \item \textbf{Format validity rate}: Among valid JSON outputs, the percentage that passed rule-based format validation.
    
    \item  \textbf{Unique sample rate}: The proportion of non-duplicate samples within the valid set, indicating diversity.

    \item \textbf{Meaning matched rate}: Among unique samples, the percentage passing semantic validation by an LLM validator, ensuring that intended actions matched actual outputs.
    \item  \textbf{Overall usability rate}: The percentage of all generated samples suitable for finetuning the target model.
\end{itemize}
\noindent 

\begin{table}[t]
\centering
\vspace{+2mm}
\caption{Synthetic data generation performance (\%) comparison among different LLMs (Since Qwen3 can switch between reasoning and non-reasoning mode, `*' indicates reasoning mode is on.)}\label{tab:data_performance}
\begin{tabular}{l | c c c c c}
\toprule
\textbf{LLM} & \textbf{JSON}  & \textbf{Format} & \textbf{Unique} & \textbf{Meaning} & \textbf{Overall}   \\
\midrule
Qwen3 (0.6B)* & 90.0 & 37.6 & 60.3 & 46.2 & 9.4 \\
Qwen3 (0.6B) & 92.0 & 46.3 & 63.6 & 50.2 & 13.6 \\
Qwen3 (8B)* & 79.0 & 18.0 & 84.8 & 93.0 & 11.2 \\
Qwen3 (8B) & 96.0 & 82.1 & 63.8 & 82.0 & 41.2 \\
Qwen3 (14B)* & 83.0 & 50.1 & 88.8 & 89.9 & 33.2 \\
Qwen3 (14B) & 78.0 & 71.8 & 90.1 & 83.2 & 42.0 \\
Gemma3 (1B) & 64.0 & 96.9 & 43.2 & 78.4 & 21.0 \\
Gemma3 (12B) & 97.0 & 82.3 & 55.6 & 83.5 & 37.1 \\
DS-R1 (1.5B) & 28.0 & 11.4 & 100.0 & 12.5 & 0.4 \\
DS-R1 (8B) & 0.0 & 0.0 & 100.0 & 0.0 & 0.0 \\
DS-R1 (14B) & 84.0 & 50.0 & 93.2 & 63.3 & 24.8 \\
Llama3.1 (8B) & 78.0 & 44.9 & 100.0 & 37.1 & 13.0 \\
Phi4 (14B) & 81.0 & 78.0 & 90.2 & 80.4 & \textbf{45.8} \\
\hline
GPT-5(nano) & 96.0 & 90.4 & 100.0 & 83.9 & \textbf{72.8} \\
\bottomrule
\end{tabular}
\vspace{-4mm}
\end{table}

\noindent As shown in Table~\ref{tab:data_performance}, some models struggled to consistently produce valid JSON, while others delivered high-quality outputs. In general, larger models achieved better performance.

\begin{table}[b]
\centering
\vspace{-4mm}
\caption{Finetuned model performance comparison among different base models (`*' indicates reasoning mode.)}\label{tab:model_performance}
\begin{tabular}{l | c c c c c}
\toprule
\textbf{Model} & \textbf{\#Parameters} & \textbf{JSON(\%)} & \textbf{Accuracy(\%)} \\
\midrule
T5 (tiny) & 15.6M & 73.42 & 45.99 \\
T5 (mini) & 31.2M & 100.0 & 85.91 \\
T5 (small) & 60.5M & 100.0 & 86.34 \\
T5 (base) & 222.9M & 100.0 & \textbf{87.53} \\
\hline
Qwen3*  & 0.6B & 99.37 & 44.34 \\
Qwen3 & 0.6B & 99.79 & 39.55 \\
Qwen3* & 8B & 99.16 & 65.27 \\
Qwen3 & 8B & 100.0 & 67.03 \\
Qwen3* & 14B & 98.95 & 68.53 \\
Qwen3 & 14B & 100.0 & \textbf{72.31} \\
Gemma3 & 1B & 99.79 & 39.79 \\
Gemma3 & 12B & 100.0 & 70.10 \\
DS-R1 & 1.5B & 68.14 & 16.49 \\
DS-R1 & 8B & 0.00 & 0.00 \\
DS-R1 & 14B & 97.89 & 67.62 \\
Llama3.1 & 8B & 83.76 & 49.14 \\
Phi4 & 14B & 95.99 & 71.47 \\
\bottomrule
\end{tabular}
\end{table}

Based on empirical results, we selected GPT-5 (nano) as the generator model for synthetic data creation. Using this dataset, we finetuned several models from the T5 family to evaluate their ability to translate natural language commands into structured JSON outputs. This comparison allowed us to identify the most efficient and accurate model for local deployment.
Accuracy was chosen as the primary evaluation metric during finetuning. It is computed through strict field-by-field matching, requiring the predicted output to exactly match the ground truth. However, this metric slightly underestimates practical performance for two reasons. First, it treats minor variations in capitalization or formatting as mismatches, even when they are semantically equivalent (e.g., ``tv\_stand'', ``TV stand'', and ``tv-stand''). Second, multiple valid interpretations of a command are possible. For example, ``create a wood chair'' could be executed in one step (create a chair with wood material) or two steps (create a chair, then change its material to wood). Although the latter is redundant, both sequences produce correct results.
The results in Table~\ref{tab:model_performance} show that finetuned T5 models, despite their smaller size, significantly outperform general-purpose LLMs on this specialized parsing task. While T5-base achieved the highest accuracy, improvements plateaued as model size increased. T5-mini, with only 31.2M parameters, reached an accuracy of 85.91\%, and doubling the size to T5-small yielded only a marginal 0.43\% gain. Given our goal of building an efficient, locally deployable tool, T5-mini offers the best trade-off between accuracy and computational cost. Therefore, we select the finetuned T5-mini as the core language model for our application.

\subsection{Demonstration of 3D Scene Generation and RRF}

\begin{figure}[t]
    \begin{subfigure}[t]{.49\linewidth}    
    \centering
    \includegraphics[width=0.9\linewidth]{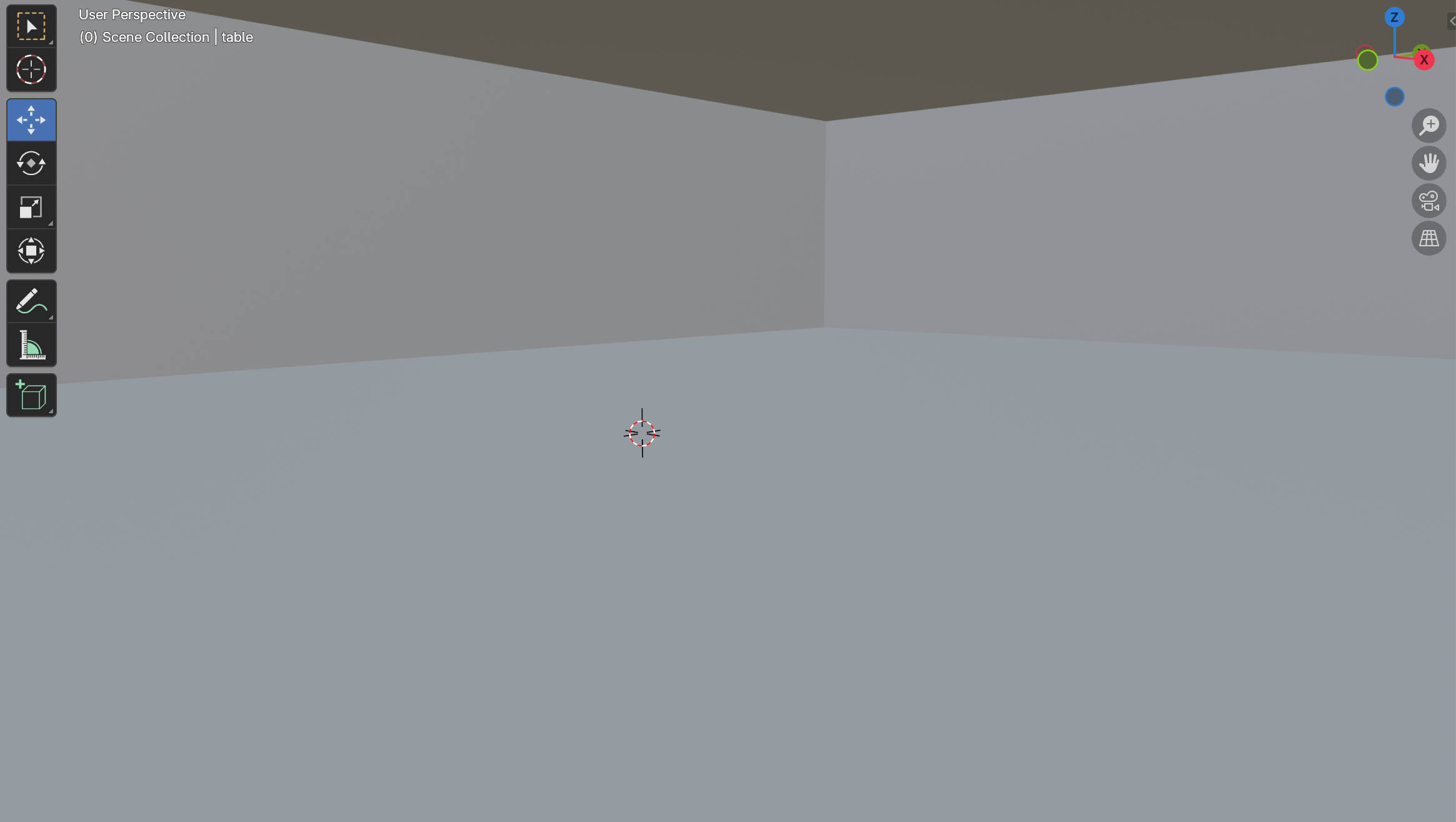}
     \caption{NIST lobby framework.}\label{fig:nist_s3_1}
    \end{subfigure}
    \hfill
    \begin{subfigure}[t]{.49\linewidth}    
    \centering
    \includegraphics[width=0.9\linewidth]{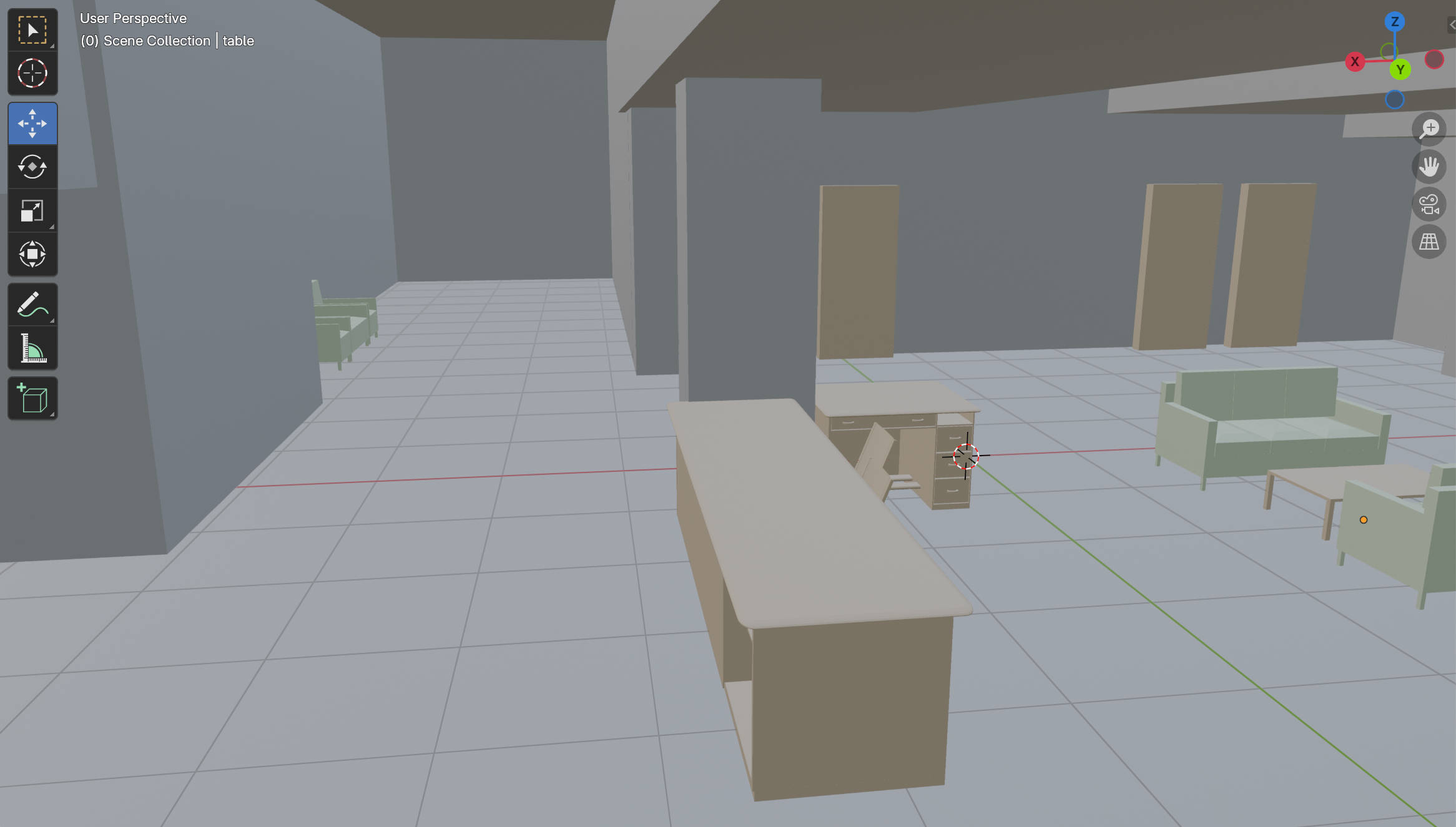}
    \caption{Furnished NIST lobby.}\label{fig:nist_s3_2}
    \end{subfigure}
\begin{subfigure}[t]{.49\linewidth}    
    \centering
    \includegraphics[width=0.9\linewidth]{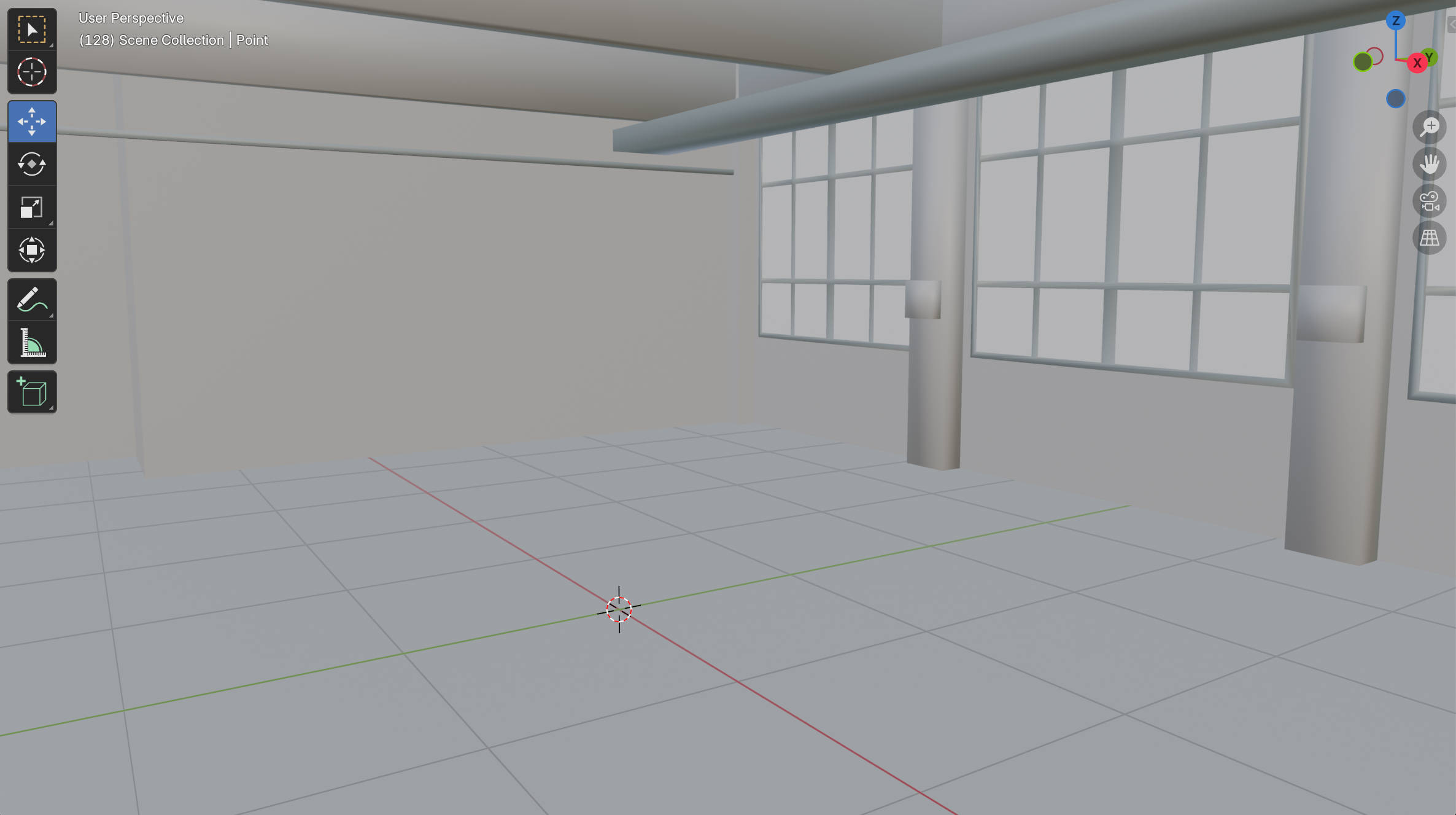}
    \caption{UW-Madison lab framework.}\label{fig:lab_1}
    \end{subfigure}
    \hfill
    \begin{subfigure}[t]{.49\linewidth}    
    \centering
    \includegraphics[width=0.9\linewidth]{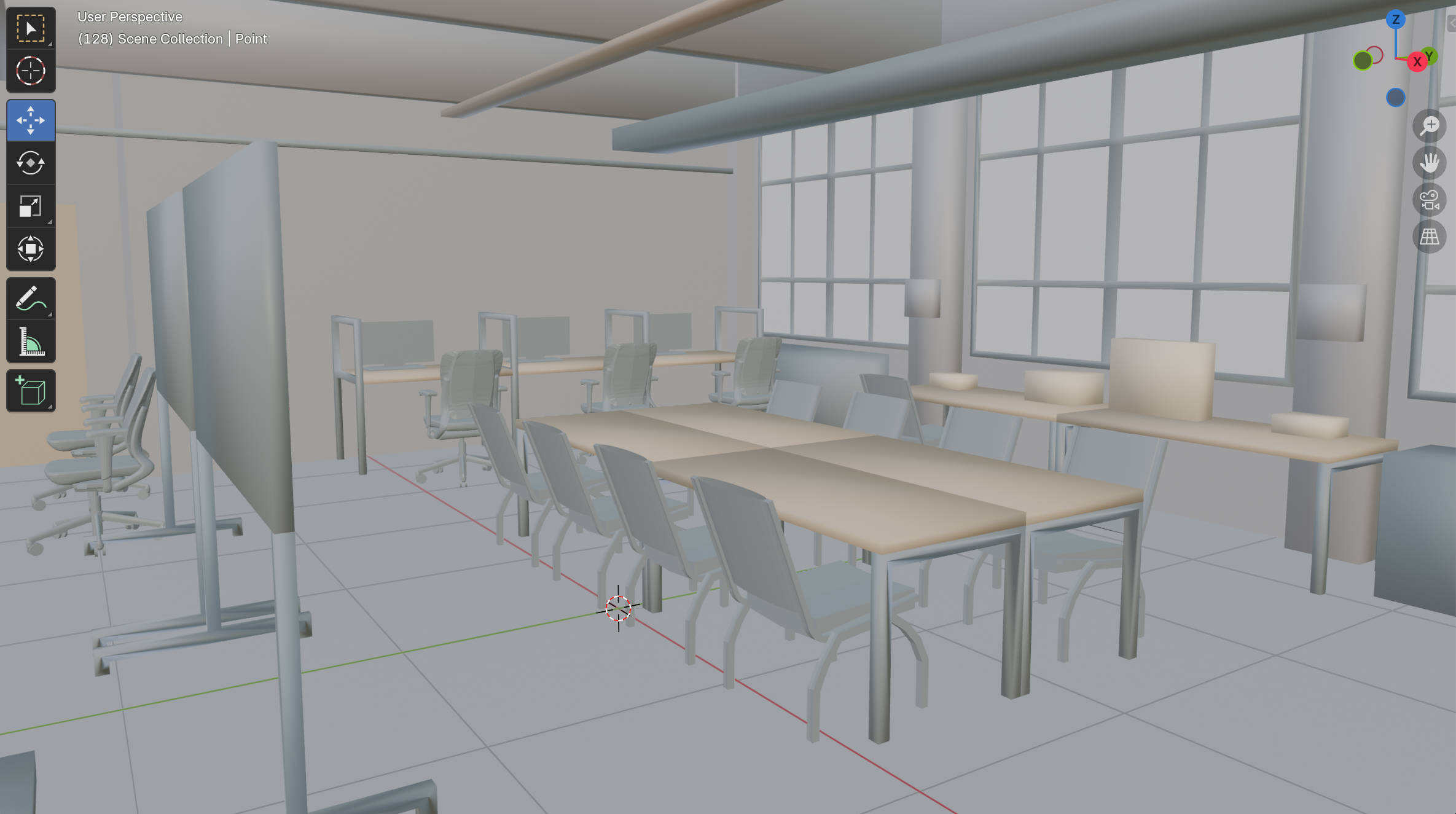}
    \caption{Furnished UW-Madison lab.}\label{fig:lab_2}
    \end{subfigure}    
     \caption{Demo of the 3D sene gereration.}
    \label{fig:3d_visual_}
    \vspace{-2mm}
\end{figure}

Using the 3D point cloud of the NIST lobby from our previous work~\cite{zhang2025rf3dgswirelesschannelmodeling}, we leveraged known room dimensions and precise furniture locations to guide scene construction. We began by creating six cubes to define the outer boundaries of the room. Once the outer shell was established, the chat-based interface was used to refine the structure by adding walls and pillars. Next, we populated the scene with furniture and other objects, specifying accurate positions and dimensions to replicate the real-world layout. Material properties were also incorporated to enable realistic radio wave propagation during RRF reconstruction. The starting frame and the furnished 3D models are shown in Figs.~\ref{fig:nist_s3_1} and~\ref{fig:nist_s3_2}, respectively.
The process for building the wireless lab at UW–Madison followed a similar approach, though without a reference point cloud. In this case, we generated a custom whiteboard model instead of using one from the built-in library and ensured that windows were accurately represented for improved RRF reconstruction fidelity. The starting frame and the completed 3D models are depicted in Figs.~\ref{fig:lab_1} and~\ref{fig:lab_2}, respectively.

\begin{figure}[t]
    \centering
    \includegraphics[width=.9\linewidth]{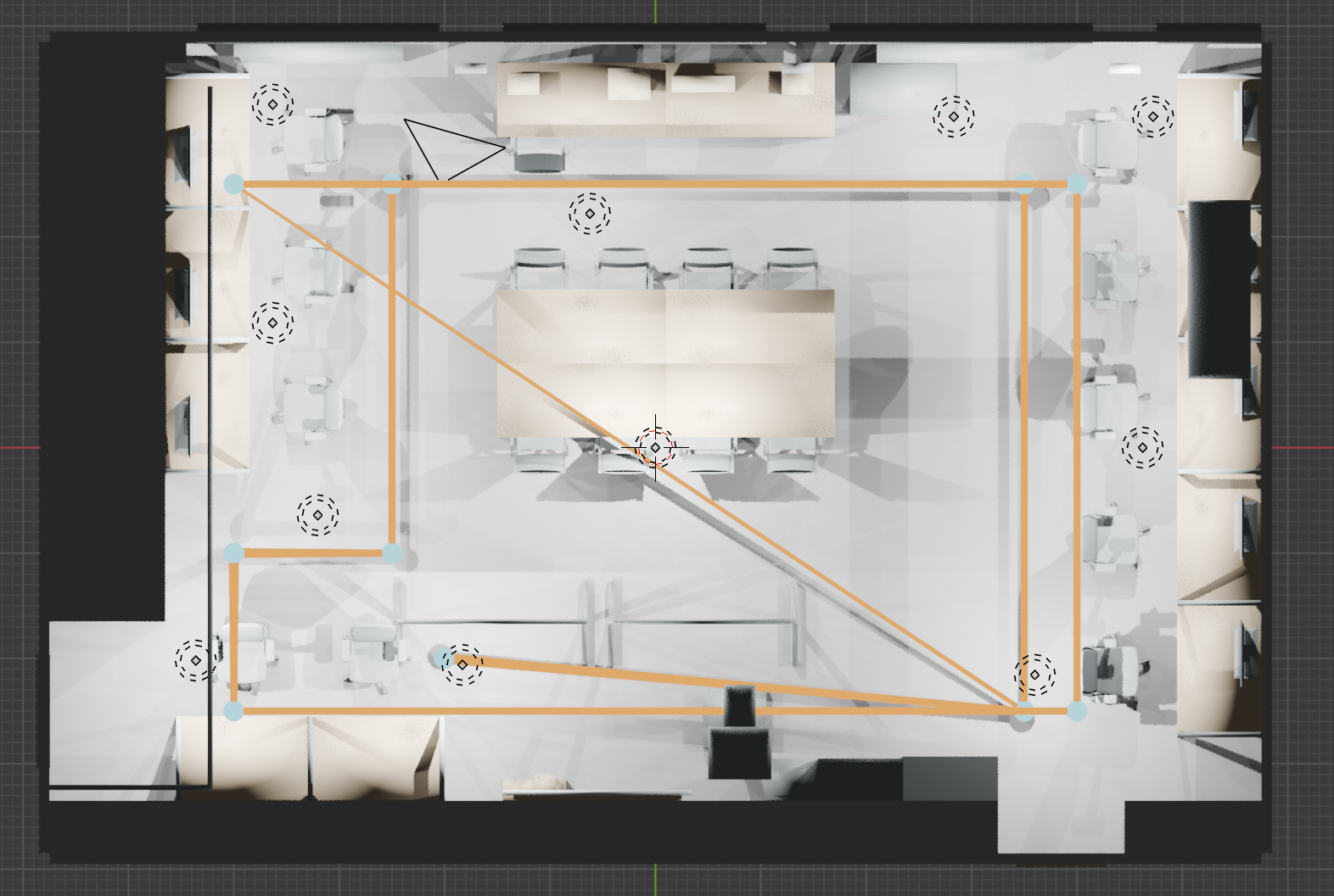}
    \caption{The top-down cross-sectional view of the UW-Madison lab scene in Blender and the configured camera trajectory (hidden ceiling and tubes).}
    \label{fig:UW-Madison_camera_track}
    \vspace{-2mm}
\end{figure}

\begin{figure}[ht!]
    \centering
    \includegraphics[width=.99\linewidth]{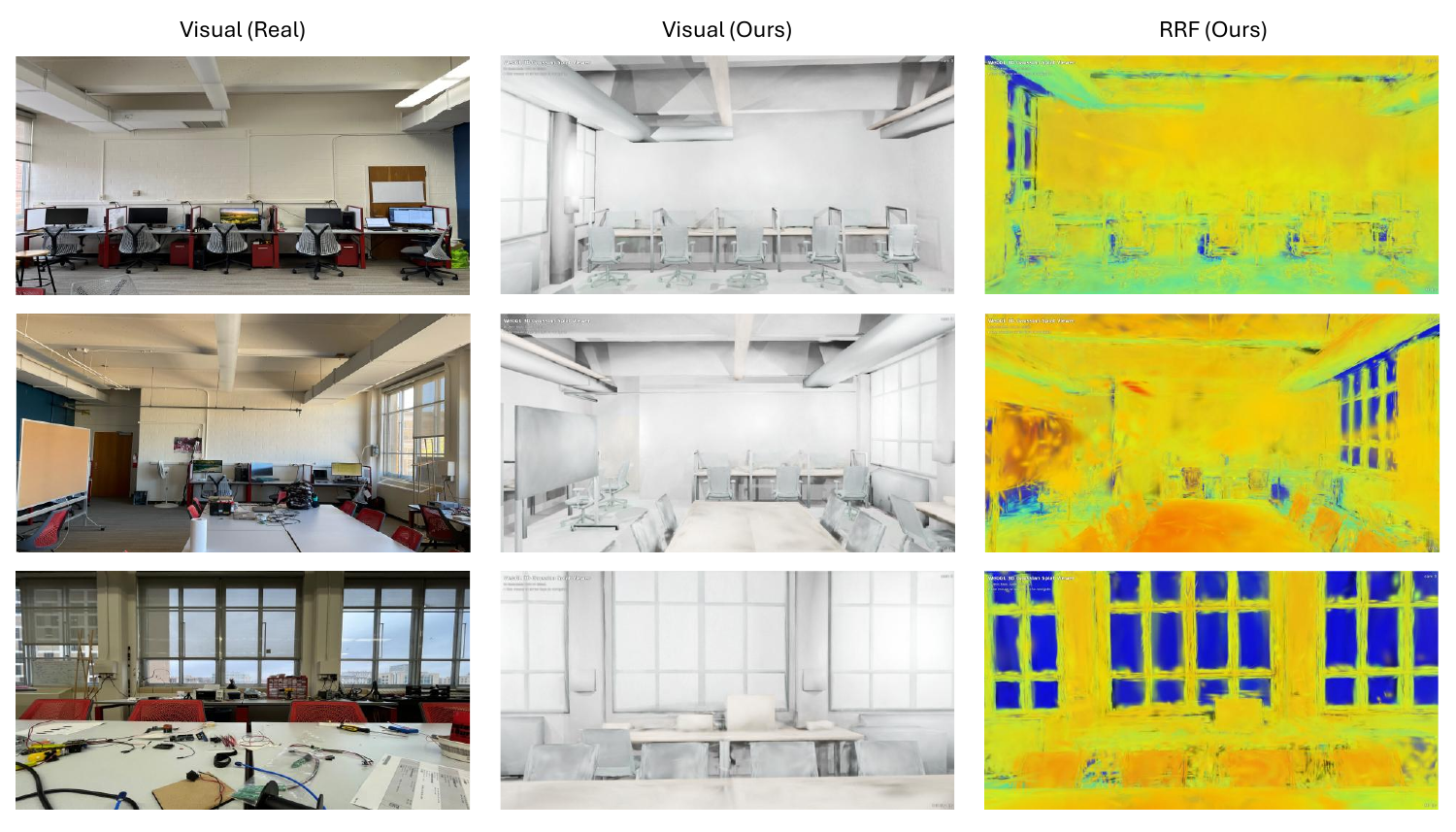}
    \caption{Demonstration of visual and RRF reconstruction of the wireless lab at UW-Madison by our method.}
    \label{fig:lab_images}
\end{figure}

\begin{figure}[ht!]
    \centering
    \includegraphics[width=.85\linewidth]{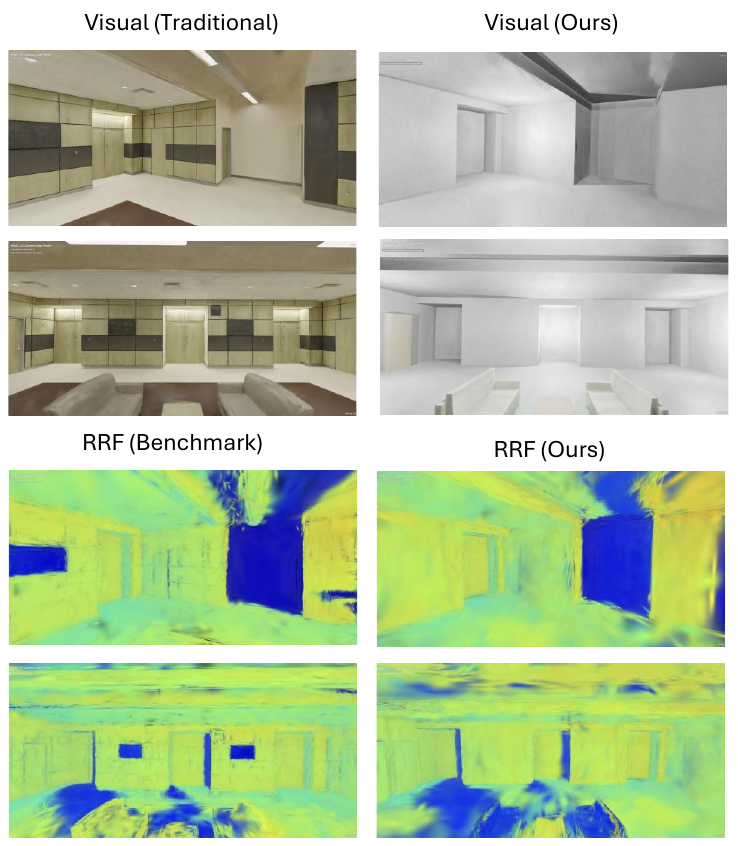}
    \caption{Comparison of visual and RRF reconstruction of the NIST lobby between traditional and our methods.}
    \label{fig:nist_images}
\end{figure}

With the constructed scenes, we implemented RRF reconstruction using the RF-3DGS pipeline. For details on the NIST lobby reconstruction, please refer to~\cite{zhang2025rf3dgswirelesschannelmodeling}. Here, we briefly outline the process for the UW-Madison lab. In Stage 1 of RF-3DGS, the standard approach was applied, with a manually designed camera trajectory to avoid blind spots. Fig.~\ref{fig:UW-Madison_camera_track} shows a top-down cross-sectional view of the lab in Blender, where the orange line represents the camera path and the light blue points indicate stop positions used to define the trajectory.
In Stage 2, we employed Blender’s BVHTree module from the mathutils library~\cite{blender_bvhtree} to detect co-located Tx and receivers Rx. A BVHTree (Bounding Volume Hierarchy Tree) was constructed from all mesh objects in the scene, enabling efficient spatial queries. Before placing a Tx or Rx, the algorithm checks whether the candidate location intersects any geometry using ray-casting and point-containment tests. This automated validation ensures that all Tx and Rx nodes are positioned in free space, preserving physical realism and preventing implausible placements. Fig.~\ref{fig:lab_images} and Fig.~\ref{fig:nist_images} illustrate the final visual and RRF reconstructions for the UW–Madison lab and NIST lobby, respectively.

Please note that the colors in the rendered models are intentionally kept plain, as they do not affect the final RRF reconstruction. The reconstructed object structures and their corresponding RRF representations closely resemble those of the actual environments.
As shown in Table~\ref{tab:performance}, both the Structural Similarity Index Measure (SSIM) and the Learned Perceptual Image Patch Similarity (LPIPS) scores are comparable to the benchmark RF-3DGS approach and outperform alternative methods.
While our proposed tool significantly streamlines the process of 3D scene generation, the current implementation is limited to a predefined set of operations (create, update, delete, etc.) and does not fully leverage the capabilities of backend AI models for generalized user inputs. Future versions will integrate more advanced LLMs, enhanced task-processing prompts, and stricter output constraints to support a broader range of natural language commands and enable more flexible scene manipulation.

\begin{table}[t]
\centering
\caption{Comparison of RRF reconstruction between the proposed method and benchmarks~\cite{zhang2025rf3dgswirelesschannelmodeling}.}\label{tab:performance}
\begin{tabular}{l | c c c c}
\toprule
\textbf{Metric} & \textbf{Ours}  & \textbf{RF-3DGS} & \textbf{NeRF$^2$} & \textbf{CGAN}   \\
\midrule
SSIM   & 0.66 &0.50 &0.46 &- \\
LPIPS  & 0.57 &0.40 &0.69 &0.87 \\
\bottomrule
\end{tabular}
\vspace{-2mm}
\end{table}

\section{Conclusion}\label{sec:conclusion}

This paper presented a locally deployable 3D modeling tool for RRF reconstruction. The tool enabled intuitive, chat-based instructions for creating and manipulating 3D objects and environments fully compatible with the RF-3DGS pipeline. Specifically, the backend integrated finetuned T5-mini and all-MiniLM-L6-v2 models alongside open-source generative frameworks such as LLaMA-Mesh and Shap-E. A chat interface facilitated user-friendly interactions, while a custom Blender executor plugin ensured seamless scene editing. Additionally, an export plugin was developed to produce 3D models compatible with RRF reconstruction requirements.
The tool was demonstrated by generating visual and RRF models of the NIST lobby and the wireless lab at UW-Madison. These case studies highlighted the system’s ability to simplify complex scene creation while maintaining compatibility with advanced wireless simulation workflows. Overall, the proposed tool provided an intuitive and efficient approach for implementing RRF, laying a strong foundation for future wireless research and spectrum planning.

\renewcommand\refname{References}
\bibliographystyle{IEEEtran}
\bibliography{Reference}

\end{document}